\documentclass[journal]{svjour3}
\usepackage{cite}
\usepackage[tight,footnotesize]{subfigure}
\usepackage{balance}
\usepackage{multirow}
\usepackage{wrapfig}
\usepackage{graphicx}
\usepackage{caption}
\usepackage[cmex10]{amsmath}
\usepackage{algorithm}
\usepackage{algorithmic}
\usepackage{url}
\usepackage[normalem]{ulem}
\usepackage{cancel}

\begin{document}
\title{Name Disambiguation from link data in a collaboration graph using temporal and topological features\footnote{A conference version 
of this paper is accepted in ASONAM 2014 conference, which is available for download from
the last author's webpage (\url{http://cs.iupui.edu/~alhasan/papers/anonymized-disambiguation.pdf}). 
The additional contributions in this journal version are explicitly
listed in the appendix section of this paper.  This research is supported by Mohammad Hasan's NSF CAREER Award (IIS-1149851).
$\dagger$ The first author and the second author contributed equally for this research
}
}
\titlerunning{Name Disambiguation from link data}

\author{Tanay Kumar Saha$^{\dagger}$, Baichuan Zhang$^{\dagger}$ and Mohammad Al Hasan}
\institute{Department of Computer and Information Science\\
Indiana University - Purdue University Indianapolis, IN 46202\\
\email{tksaha@cs.iupui.edu, bz3@umail.iu.edu, alhasan@cs.iupui.edu}}

\maketitle

\begin{abstract}
In a social community, multiple persons may share the same name, phone number or
some other identifying attributes. This, along with other phenomena, such as
name abbreviation, name misspelling, and human error leads to erroneous
aggregation of records of multiple persons under a single reference. Such
mistakes affect the performance of document retrieval, web search, database
integration, and more importantly, improper attribution of credit (or blame).
The task of entity disambiguation partitions the records belonging to multiple
persons with the objective that each decomposed partition is composed of records
of a unique person.  Existing solutions to this task use either biographical
attributes, or auxiliary features that are collected from external sources,
such as Wikipedia.  However, for many scenarios, such auxiliary features are
not available, or they are costly to obtain. Besides, the attempt of collecting
biographical or external data sustains the risk of privacy violation. In
this work, we propose a method for solving entity disambiguation task from link
information obtained from a collaboration network. Our method is non-intrusive
of privacy as it uses only the time-stamped graph topology of an anonymized network.
Experimental results on two real-life academic collaboration networks show that
the proposed method has satisfactory performance. 
\end{abstract}

\section{Introduction}
On January 17, 2014, in his speech regarding the usages of phone surveillance
data by NSA (National Security Agency), the USA President Barack Obama said,
``This program does not involve the content of phone calls, or the names of
people making calls. Instead, it provides a record of phone numbers and the
times and lengths of calls---metadata that can be queried if and when we have a
reasonable suspicion that a particular number is linked to a terrorist
organization.'' In this talk he also mentioned the importance of balancing
security and privacy in all surveillance works of government agencies. However,
making this balance is not an easy task; respecting privacy does not allow
tapping into someone's non-public biographical records; on the other hand,
constrained analysis without detailed biographical data leads to
numerous false identification and entity mixup. In this work, we are concerned
with solving the task of entity disambiguation without using biographical
information---the input to our solution is link data collected from anonymized
collaboration networks, similar to the one that Mr. Obama has explained.

Entity disambiguation~\cite{Cucerzan.Silviu:07,Sen:12} is
not a new problem. In fact, named entity disambiguation has been a long
standing problem in the field of bibliometrics and library science. The key
reason for this is that many distinct authors share the same name, specifically
considering the fact that the first name of an author is typically written in
abbreviated form in the citation of many scientific journals. Thus,
bibliographic servers that maintain such data may mistakenly aggregate the
records from different persons into a single entity.  For example consider
DBLP, the largest bibliographic website of computer science. In DBLP, there are
at least $8$ distinct persons named Rakesh Kumar, and their publications are
mixed in the retrieved citations.  The ambiguity in Chinese names is more
severe, as many Chinese share a few family names such as Wang, Li, and Zhang.
An extreme example is Wei Wang. According to our labeling, it corresponds to
over $200$ distinct authors in DBLP! To correctly assess the impact of a
researcher in a research field, correct attribution of research works is
essential, so entity disambiguation has been extensively addressed by
researchers in information retrieval and data mining. Note that, a
related problem considers the task of merging multiple name references into a single
entity, where the records belonging to a single person has been erroneously
partitioned into multiple name references ~\cite{Bhattacharya.Getoor:04, 
Bhattacharya.Getoor:06,Whang.Hector.ea:10,Sarawagi.Bhamidipaty:02, Whang.Hector.ea:09}.
This task is more popularly known
as entity deduplication or record linkage, and it is not the focus of this work.

Many research works are proposed to solve the entity disambiguation.  Among
those, some are specifically targeted for solving the name entity
disambiguation~\cite{Bunescu.Pasca:06,Cucerzan.Silviu:07,Han.Giles.ea:04,Han.Zha.ea:05,Einat.Cohen.ea:06}.
Existing works mostly use biographical features, such as name, address,
institutional affiliation, email address, and homepage; contextual features,
such as coauthor/collaborator, and research keywords; and external data such as
Wikipedia~\cite{Cucerzan.Silviu:07}. From methodological point of view, some of
the works follow a supervised learning
approach~\cite{Han.Giles.ea:04,Hermansson.Kerola.ea:13}, while others use
unsupervised
clustering~\cite{Han.Zha.ea:05,Cen.Dragut.ea:13,Wang.Li.ea:08,Malin:05}.  There
exist quite a few solutions that use graphical
models~\cite{Zhang.Tang.ea:07,Tang.Fong.ea:12,Wang.Tang.ea:11,Bhattacharya.Getoor:06}.
What is common among all these works  is that they use many biographical features
including name, and affiliation, so they cannot protect the privacy of the
actors in the dataset. Using biographical features is acceptable for entity
disambiguation of authors in the field of bibliometrics, but it raises a
serious concern when it is being used for applications related to national
security; in such an application, it is more desirable to use a pre-filter that
identifies a small list of suspicious data references for which biographical
data can be queried after the approval of a privacy management officer. Besides
the concern of privacy, significant cost is also involved in collecting
biographical information, which impedes the effective utilization of the existing
methodologies for solving entity disambiguation.

In this work, we consider an anonymized social network. Each node in this
network corresponds to a reference to a name entity, and each edge corresponds
to collaboration among different name entities. The edges are labeled with
time-stamps representing the time when a collaboration took place. As we have
discussed earlier, we can think of such a network as the anonymized email/communication
network that the NSA uses to identify suspects. 
Our solution to entity disambiguation in an anonymized network uses
the timestamped network topology around a vertex of the network and by using an
unsupervised method it produces a real-valued score for that vertex. This score
represents the degree to which a given anonymized reference (a vertex) is
pure. The smaller the score, the more likely that the reference may comprise of
records of multiple real-life entities. For a given vertex, the method provides
the desired score in
a few seconds, so one can always use it as a pre-filter to identify a small set
of target nodes for which more thorough analysis can be made subsequently. Alternative
to an unsupervised approach, our method can also be adapted to a supervised 
classification system for predicting the purity status of a node,
when a labeled dataset is available.\\

We claim the following contributions in this work:

\begin{itemize}
\item We design the task of solving name entity disambiguation 
using only graph topological information. This work is motivated
by the growing need of data analysis without violating the privacy
of the actors in a social network. 

\item We propose a simple solution that is robust and it takes only a few seconds 
to disambiguate a given node in real-life academic collaboration networks. The proposed method
returns a real-valued score to rank the vertices of a network based on
their likelihood for being an ambiguous node. So the score
can be used as a pre-filter for identifying a small set of ambiguous references
for subsequent analysis with a full set of features. Besides, the
score can also be used independently as a feature for classification
based solutions for entity disambiguation.

\item We use two real-life datasets for evaluating the performance of our
solution. The results show that the method performs satisfactorily, considering
the fact that it uses only the topology of a node in its analysis.
\end{itemize}


\section{Related Work}

In existing works, name disambiguation task is studied for various entities;
examples include disambiguation on Encyclopedic knowledge or Wikipedia
Data~\cite{Bunescu.Pasca:06,Cucerzan.Silviu:07,Kataria.Kumar.ea:11}, citation
data~\cite{Han.Giles.ea:04,Han.Zha.ea:05,Tan.Kan.ea:06,Yin.Han.ea:07}, email
data~\cite{Einat.Cohen.ea:06}, and text documents~\cite{Sen:12,Li.Wang.ea:13}.

In terms of methodologies, both
supervised~\cite{Han.Giles.ea:04,Bunescu.Pasca:06} and
unsupervised~\cite{Han.Zha.ea:05} approaches are considered. For supervised
method, a distinct entity can be considered as a class, and the objective is to
classify each event to one of the classes. Han et al.\cite{Han.Zha.ea:05} use
such a framework, and propose two supervised methods, one using a generative
model, and the other using SVM. In another supervised approach, Bunescu et al.~\cite{Bunescu.Pasca:06} 
solve name disambiguation by designing and training a
disambiguation SVM kernel that exploits the high coverage and rich structure of
the knowledge encoded in an online encyclopedia. However, the main drawback of
supervised methods is the lack of scalability, and the unavailability of labeled
data for training. It is also impractical to train thousands of models, one for
each individual, in a large digital library.

For unsupervised name disambiguation, the
collaboration events are partitioned into several clusters with a goal that
each cluster contains the events corresponding to a unique entity. Han et
al.\cite{Han.Zha.ea:05} propose one of the earliest unsupervised name disambiguation 
methods, which is based on K-way
spectral clustering. They apply their method for name disambiguation in an academic 
citation network. For each name dataset, they calculate a Gram matrix representing similarities 
between different citations and apply
$K$-way spectral clustering algorithm on the Gram matrix to obtain the desired
clusters of the citations. In another unsupervised approach, Cen et al.\cite{Cen.Dragut.ea:13} compute
pairwise similarity for publication events that share the same author name string
(ANS) and then use a novel hierarchical agglomerative clustering with adaptive
stopping criterion (HACASC) to partition the publications in different author
clusters. Malin~\cite{Malin:05} proposes another cluster-based method that uses
social network structure.

Probabilistic relational models, specifically graphical models have also been
used for solving entity disambiguation task. For example, authors
in~\cite{Zhang.Tang.ea:07} propose a constraint-based probabilistic model for
semi-supervised name disambiguation using hidden Markov random fields (HMRF).
They define six types of constraints and employ EM algorithm to learn the HRMF
model parameters.  In another work, Tang et
al.~\cite{Tang.Fong.ea:12,Wang.Tang.ea:11} present two name disambiguation methods that are based
on pairwise factor graph model.  They target name disambiguation in
academic datasets. In their work, the authorship of a paper is modeled as edges between
observation variables (papers) and hidden variables (author labels).  Features
of each paper and relationships, such as, co-publication-venue and co-author,
have impact on the probability of each assignment of labels. The similarity
between two clusters is encoded in different factors (edge potentials) on
different features. The clustering process iterates over different author label
assignments and selects the one with maximal probability.  An improved version
of the above work uses user feedback and is being used in the real-life
Arnetminer system (\url{arnetminer.org}) for disambiguation~\cite{Wang.Tang.ea:11}.  LDA based
context-aware topic models has also been used for entity
disambiguation~\cite{Sen:12}.

To build the features for classification, clustering, or probabilistic models,
most of the existing works use biographical and contextual attributes of the
entities or external sources such as Wikipedia, Web search results and online
encyclopedia. In almost every work on name disambiguation, person's name, email
address, and institutional affiliation are used. It is not surprising, because
biographical features are highly effective for name disambiguation. For
instance, a set of recent works~\cite{Chin.Juan.ea:13,Liu.Lei.ea:13} report
around 99\% accuracy on a data mining challenge dataset prepared by Microsoft
research.  These works use supervised setup with many biographical features;
for instance, one of the above works even predict whether the author is Chinese
or not, so that more customized model can be applied for these cases.

In this work, we consider the task of name disambiguation using
only the network topological features in an anonymized collaboration graph.
We found a recent work~\cite{Hermansson.Kerola.ea:13} which also
has a similar objective. They consider this problem in the supervised setting
where they are provided with a base graph and a set of nodes labeled as
ambiguous or unambiguous. They characterize the similarity between two nodes
based on their local neighborhood structure using graph kernels and solve the
resulting classification task using SVM. We will show in the experiment section
that our method performs significantly better than this method in terms of both
speed and accuracy. Note that, a preliminary version of our paper
is already published as a short conference article~\cite{Zhang.Saha.ea:14}. 

\section{Solution Overview}
We assume that a collaboration network $G(V, E)$ is given, where each 
node $u \in V$ represents an entity reference which in real-life may
be linked to multiple persons. For every edge $e\in E$ we are also provided
with a list $T(e)$ which represents the discrete time-point at which the 
collaboration events between the corresponding nodes have taken place.
Our objective is to predict how likely it is that the node $u$ is a 
multi-node, i.e., it comprises of collaboration records of multiple persons.
We use a linear model to produce a numeric score $s(u)$, which represents the
likelihood that $u$ is a pure node.

To solve the problem in the setup discussed in the last paragraph, we first
construct the ego network of $u$, $G_u \subseteq G$, which is an induced
subgraph of $G$ consisting of ego node $u$ and all of its direct neighbors
(these nodes are called ``alters''). Since $G_u$ is an induced subgraph, it
preserves the ties between $u$ and the alters and also the ties between a pair
of alters. We hypothesize that if $u$ is a multi-node, the graph $G_u$ will
form many disjoint clusters, once the node $u$ and all of its incident edges
from $G_u$ are removed; each of these clusters corresponds to one of the many
real-life entities that have been merged together under the reference $u$.
This hypothesis is built from the transitivity property of a social network,
which states that the friends of your friend have high likelihood to be friends
themselves~\cite{Mathew:08}. Thus, if $v$ and $w$ are friends of $u$, with high
likelihood, there are edges between $v$ and $w$. However, when $u$ is a
multi-node corresponding to $k$ different people, the friends of $u$ are
partitioned into at least $k$ disjoint clusters.

\begin{figure}[h]
\centering
\includegraphics[width=0.35\textwidth] {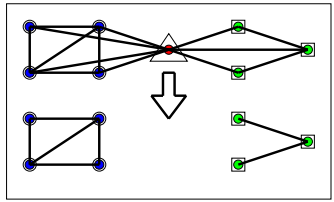}
\caption{A toy example of clustering based entity disambiguation}
\label{fig:graph}
\end{figure}

In Figure~\ref{fig:graph}, we illustrate our hypothesis. Assume that the
triangle shaped node is $u$, and the graph in this figure corresponds to the
ego network of $u$, $G_u$. We also assume that $u$ is a multi-node
consisting of two name entities. So the removal of the node $u$ (along with all of
its incident edges) from $G_u$ makes two disjoint clusters; this phenomenon is
illustrated in the lower part of figure~\ref{fig:graph}. The vertices of these 
clusters are shown using circles and squares respectively. 

However, there are several caveats in the above simplified formulation.
Particularly, the above formulation may yield numerous false alarms for various
reasons even if a node $u$ is not a multi-node. First, if the name entity at
$u$ participates in multiple communities, her neighbors may form disjoint
clusters. Second, the neighborhood of the entity $u$ may have several distinct
clusters considering the temporal axis, which happens if $u$ changes job,
institution or affiliation. False negatives also occur, though with a lesser
likelihood. For a multi-node $u$, if the significant parts of the collaboration 
activities of $u$ comprise of only one name entity, the remaining weaker entities 
under $u$ contribute poorly in the score $s(u)$, which may prevent from categorizing 
$u$ as a multi-node. Another challenge is that due to the power-law behavior
of a typical collaboration graph $G$, the neighborhood graph $G_u$ has varying
size and density, which affects the comparison of the score value $s(\cdot)$ of 
various nodes in $G$. We take into consideration each of these problems
in our proposed solution, as explained in next sections.

\section{Methods}

We assume that a collaboration network $G = (V,E)$ and a ego node $u$ is given,
where $u \in V$ represents an entity reference, which in a real-life scenario may 
be linked to multiple persons. From $G$ and $u$, we construct the ego network
of $u$, $G_u = (V_u, E_u) \subseteq G$. For each edge in $G_u$, say $e_u = (v,w) \in E_u$,
$T(e_u)$ represents the set of collaboration events between $v$ and $w$ that are
captured by $G_u$; i.\ e.\,
$T(e_u) = \left\{\langle n_{i}, t_{i}\rangle\right\}_{1\le i\le \mid T \mid}$, here,
$n_{i}$ is the number of collaboration events at time $t_i$. From $T(e_u)$, we compute
a similarity value between $v$ and $w$ under $G_u$ using an exponential decay function as 
shown below:
\begin{equation*}
\label{eq:exponential-decay}
W_{G_u}(v, w)=\displaystyle\sum_{i=1}^{|T|}{n_i\times\exp{\frac{-(t_{\max}-t_i)}{\tau}}}
\end{equation*}
In the above equation, $t_{max}$ denotes the most recent time when a collaboration
event happened between any two vertices in $G_u$. $\tau$ is a tuning parameter that 
one can use to control the rate of the decay. On academic collaboration networks, 
for which the time unit is year, we use $\tau$ ranging between 5 to 10. Empirical 
results on such networks also show that the performance of the system is not sensitive 
to the choice of $\tau$.  The above definition of edge similarity function 
rewards more weight to recent collaboration events than to older collaboration events.\\

\noindent {\bf Example:}
Given $G_u = (V_u, E_u)$;  $v, w \in V_u$. Let us assume 
$t_{max} = 2014$. Besides, $v$ and $w$ has $2$ collaboration events in $2014$, $3$ collaboration 
events in $2013$ and 
$4$ collaboration events in $2010$; thus, $T((v,w)) = \{\langle 2, 2014\rangle, 
\langle 3, 2013\rangle, \langle 4, 2010\rangle\}$.
By setting $\tau=5$, we get,
$W_{G_u}(v, w) = 2 \times \exp{\frac{-(2014-2014)}{5}} + 3 \times \exp{\frac{-(2014-2013)}{5}} 
+ 4 \times \exp{\frac{-(2014-2010)}{5}} \approx 6.25$.~\qed \\

\subsection{Obtaining Cluster Quality Score}\label{sec:tm}
Given the collaboration network, $G$ and a specific vertex $u$, the next step of
our method is to construct $G_u$---the ego network of $u$.  Then, we cluster the graph $G_u \setminus \{u\}$ using a graph
clustering algorithm. The objective of this clustering is to group $u$'s
neighbors in different clusters such that the cross-edges between different
clusters are minimized. The reason of removing $u$ is to find whether the
neighbors of $u$ are strongly connected by themselves without using $u$ as an
intermediate vertex. 

We utilize Markov Clustering (MCL) for the clustering 
task. MCL uses the graph's natural transition probability matrix 
to cluster a graph by combining random walks with two alternating operations 
(expansion and inflation)~\cite{Dongen.Stijn:08}. 
There are several reasons for the choice of MCL. First, MCL is one of the fastest
graph clustering methods; competing other methods, such as, spectral clustering and
its variants~\cite{Luxburg:07} compute eigenvectors of graph Laplacian, which could be costly
for large graphs. For our work, we found that for a graph with 
several thousands vertices, MCL finishes with good clustering results only within
a few seconds. Second, MCL does not require the number of clusters as one
of the input parameters, which works well for our setting as we have no information
regarding the number of communities in which the node $u$ participates. Finally,
MCL is robust against the choice of parameters. It has only one parameter,
called {\em inflation}, which we set to the default value in all of our experiments.

\subsubsection{Normalized Cut based Score}
For our task, we are mainly interested in obtaining a score reflecting the quality
of clustering. There are various evaluation criteria for clustering, including
ratio cut, normalized cut~\cite{Luxburg:07} and modularity~\cite{Newman:06}. Among these,
we choose the normalized cut based clustering score as it reflects the ratio
of the similarity weight-sum of the inter-cluster edges and the same for all the edges
in the graph. The equation of normalized cut score for a node $u$ is shown below:

\begin{equation}
\label{eq:normalized-cut}
NC = \displaystyle \sum_{i=1}^{k} \frac{ W(C_i, \overline{C_i})}{W(C_i, C_i)+W(C_i,\overline{C_i})}
\end{equation}
where $W(C_i,C_i)$ denotes the sum of weights of all internal edges and
$W(C_i,\overline{C_i})$ 
is the sum of weights for all the external edges and $k$ is the number of clusters 
in the graph. We compute $NC$ value by independent calculation after we obtain 
the clusters of $G_u$ using MCL algorithm.

For a node $u$, the normalized cut (NC) score denotes the clustering tendency
of the neighbors of $u$. If the value is high, then the clustering tendency
of $u$'s neighbors is poor, so $u$ is less likely to be a multi-node. On the other
hand, if the value is small, then $u$' neighbors are well clustered and $u$ has
a high probability to be a multi-node.
For example, in the bibliographic domain, if a multi-node $u$ in a
co-authorship network represents two researchers who share the same name, with
a high likelihood, they will have a distinct set of co-authors. After clustering
the graph $G_u \setminus \{u\}$ using MCL, we expect to obtain two dominant clusters
that are disconnected (or very sparsely connected),
each representing the set of co-authors of each of the two researchers.
One problem with the $NC$-score is that it is not size invariant, i. e., if a
node $u$ has many clusters, its NC-score is large, so this score is not
that useful when we compare the $NC$-scores of many nodes to find the one which
is likely to be a multi-node. In order to address this issue, we normalize the score 
by the number of clusters as shown below:
\begin{equation}
\label{eq:nc-score}
NC\mbox{-}score=\frac{NC}{k}
\end{equation}
where $k$ is the total number of clusters that we obtain using MCL, 
given the ego network of $u$, $G_u$.

\subsection{Obtaining Temporal Mobility Score}
NC-score is a good metric to represent the degree at which a given
anonymized entity is pure. However, for many real-life datasets, this score
yields many false positives. Assume, a vertex in a social network represents one
real-life entity, but its collaboration network evolves over time because of
entity mobility. For such a vertex, the $NC$-score is small due to disjoint
clusters of neighbors that are formed as the entity moves along with the time;
this leads to a false positive prediction that the entity is likely to be a
multi-node.  Since the anonymized collaboration data has the time stamp of
the collaboration event, we use this information to obtain a second score that
we call Temporal Mobility (TM) score. This score indicates how likely is that
the specific entity has moved along with the time. Note that, temporal mobility
is not a new concept, it has been studied by social scientists earlier;
for instance, there are works to understand the temporal mobility of academic
scientists as they change their jobs~\cite{Allison.Long:87}.

To compute the likelihood that a given vertex, $u$, has experienced temporal
mobility, we obtain the $TM$-score (temporal mobility score) of $u$. This score
reflects the extent by which the node $u$ collaborates with different neighbor
clusters (obtained using MCL) at a distinct time range. Below we discuss the
process that computes the $TM$-score of a node $u$.

Let's assume that $C_{i:1\le i \le k}$ are $k$ different clusters that we
obtain from MCL during the cluster quality computation stage. To model $u$'s
collaboration with a cluster $C_i$, we first obtain a vector, $Z_i$ of size
$|T|$ (the number of distinct time intervals in $u$'s collaboration history),
in which each entry denotes the total number of collaboration events between
$u$ and the members of $C_i$ at that specific time interval. If a collaboration
event with $u$ consists of $l (\ge 2)$ actors (besides, $u$), each of the actors in
cluster $C_i$ contributes $1/l$ to the appropriate entry in $Z_i$. Say, $u$
publishes a paper with $l$ co-authors at year $t$. After clustering, $m$ of the
co-authors out of the $l$ co-authors of that paper belong to the cluster $C_i$,
this event additively contributes to $Z_i$ vector's $t$'th position by
$m/l$.  Thus for a multi-party events, a collaboration event with $u$ can be
distributed among different clusters, in case $u$'s accomplices for this event
are distributed among different clusters. Thus, by iterating over all the collaboration
events of $u$, we compute $k$ different vectors, $Z_i:{1 \le i \le k}$, each
such vector corresponds to $u$'s collaboration with the entities in one of the 
clusters, $C_i$. Finally, we use centered moving average to smooth the
$Z_i$ vectors.
\begin{figure}[h!]
\centering
\includegraphics[scale=0.30,angle=-90] {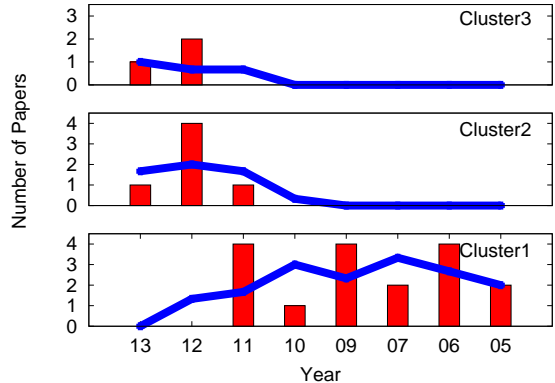}
\caption{Temporal mobility example}
\label{fig:hist-cluster1}
\end{figure}

For the purpose of illustration, we present an example of temporal mobility of an
entity in Figure~\ref{fig:hist-cluster1}. In this Figure we can observe 3 histograms;
each histogram shows the yearly count of collaboration events that this entity
has with other entities in one of his neighbor-clusters. We also present the same
information after smoothing with the centered moving average.  We draw the
histograms in this figure using real-life data of a researcher in DBLP academic
collaboration network. In the DBLP data, all the authorship records of this
researcher point to a single name reference.  However, when we use MCL
clustering, we obtain 3 dominant clusters for this person, with almost no
inter-cluster edges leading to a small $NC$-score.  This suggests that this
entity has a high chance to be a multi-node. But, as shown in
Figure~\ref{fig:hist-cluster1}, the association pattern of this entity with the
three different clusters suggests temporal mobility. During 2005-2010 time
period, the entity has almost dedicated association with the entities in
Cluster 1; which is followed by a divided association among the entities in
Cluster 2 and Cluster 3 for the time period 2011-2013.

The name of the entity that we discuss here is Dr. Honglak Lee. The first
cluster corresponds to his collaboration with his co-authors when he was a PhD
student in the Stanford University. The second cluster denotes his
collaboration with his PhD students at the University of Michigan.  The third
cluster represents his collaboration with his colleagues in the same
university. This example illustrates that how temporal mobility of an entity
can lead to a false positive multi-node when only NC-score is used for this
qualification.

\subsubsection{Temporal Mobility score using KL Divergence}

Our main objective in this step is to obtain the Temporal Mobility (TM) score
of a node $u$ after we obtain $u's$ cluster-wise collaboration vectors, $Z_i$. 
For this we convert each of the smoothed $Z_i$ vectors to a
discrete probability distribution by normalizing these vectors appropriately.
Thus, each of $u$'s neighbor-clusters are represented by a discrete probability
vector. To compute the temporal divergence of these clusters we use
Kullback-Leiber (KL) divergence which is a measure
of divergence between two probability distributions. $D(P \parallel Q)$ denotes
the KL divergence between two probability distributions $P$, $Q$, and it is
defined on a finite set $\chi$ as below:
\begin{equation*}
\label{eq:kl-divergence}
D(P \parallel Q)=\displaystyle \sum_{x \in \chi}P(x)\log\frac{P(x)}{Q(x)}
\end{equation*}
This value is large, if the two distributions are different, and vice-versa. 
Note that, $D(P \parallel Q)$ is an asymmetric metric. So for our task we use symmetric KL divergence
which for the distributions $P$ and $Q$ are $D(P \parallel Q)+D(Q \parallel P)$.
Also, to avoid the scenario when the discrete distributions of $P$ and $Q$ contain a zero element,
we adopt Laplace correction that assigns a small probability (0.01)
to those entries.

Now, $TM$-score of $u$ is simply the weighted average value of the symmetric KL
divergence of all pairs of $Z_i$ (normalized to 1) vectors. For each pair,
$Z_i$, and $Z_j$, we first compute $D(Z_i \parallel Z_j)$ and $D(Z_j \parallel
Z_i)$. The weight of the divergence between $Z_i$ and $Z_j$, is denoted as $w(Z_i, Z_j)$; 
we use the sum of the number of events in the cluster $C_i$ and $C_j$ as
this weight. Using such weighting, the KL-divergence between dominant clusters
contribute more in the $TM$-score. Similar to the case of $NC$-score, we also
normalize $TM$-score by the number of clusters, so that different nodes with
diverse number of clusters can be compared. The overall computation can be
shown using the following equation:
\begin{equation}
\small
TM\mbox{-}score =\frac{\displaystyle \sum_{i=1}^{k-1} \sum_{j=i+1}^{k} w(Z_i,Z_j) \cdot 
\bigg(D (Z_i \parallel Z_j)+ D(Z_j \parallel Z_i)\bigg)} 
{ k \times  \sum_{i=1}^{k-1} \sum_{j=i+1}^{k} w(Z_i, Z_j) }
\label{eq:tm-score}
\end{equation}
The higher the value of $TM$-score of an entity, the higher the likelihood that
the node is not a multi-node. Rather it has experienced the temporal mobility
phenomenon along its overall time intervals.

\subsection{Linear Model using $NC$-score and $TM$-Score}
We can use $NC$-score and $TM$-score for unsupervised learning.
For an unsupervised case, we simply predict a score $s(u)$ for a node $u$. The
higher the score, the more likely that the node is a pure node. For this we use
a linear model with only one model parameter $\alpha$, which is positive, because
both $NC$-score and $TM$-score have larger values for a pure node and smaller value
for a multi-node.  Thus, the score $s(u)$ of a node $u$ is simply:
\begin{equation}
s(u)= NC\mbox{-}score(u) + \alpha \cdot TM\mbox{-}score(u)
\label{eq:s-score}
\end{equation}
The model parameter $\alpha$ should be set manually depending on the nature of the dataset. 
In co-authorship networks, a small $\alpha$ value in the range from 0.1 to 0.2 works well.
The benefit of this unsupervised method is that we can simply work on a small
collection of nodes independently. By sorting the $s$-score of those nodes in increasing
order, we can identify the top-ranked nodes that are likely to be suspected of being
a multi-node.

\begin{algorithm}
\renewcommand{\algorithmicrequire}{\textbf{Input:}}
\renewcommand{\algorithmicensure}{\textbf{Output:}}
\caption{\bf Unsupervised-Disambiguation$(G,u)$}
\label{alg1}
\begin{algorithmic}[1]
\REQUIRE $G$, $u$
\ENSURE $s(u)$
\STATE Construct the ego network $u$, $G_u$ using the similarity weight between vertices
\STATE Remove $u$ from $G_u$ and apply MCL to get $k$ clusters $\{C_i\}_{1\le i \le k}$ 
\STATE Using Equation ~\ref{eq:normalized-cut} and Equation ~\ref{eq:nc-score}, compute $NC$-score
\STATE For each cluster $C_i$,compute normalized $Z_i$ vector using timestamp of association
\STATE Use Equation ~\ref{eq:tm-score} to compute $TM$-score
\STATE Using a set of validation dataset, tune the $\alpha$ value for the given dataset
\STATE \textbf{return} $s(u) = NC\mbox{-}score(u) + \alpha \cdot TM\mbox{-}score(u)$
\end{algorithmic}
\end{algorithm}

\subsection{\textbf{Pseudo-code and Complexity Analysis}}

The pseudo-code of the entire process for the unsupervised setup is given in
Algorithm~\ref{alg1}. It takes an input graph $G$ and a specific ego node $u$ as
input and generates the numeric score of $u$, $s(u)$ as output.
Line $1$ computes the similarity weights for the edges of the ego network of $u$, 
$G_u$. Line 2 removes $u$ from $G_u$, and applies MCL clustering
method to cluster the similarity graph $G_u$. We assume that this clustering
yields $k$ clusters, $\{C_i\}_{1\le i \le k}$. From these clusters, Line 3 obtains
the normalized cut based score. For each cluster (say,
$C_i$), Line 4 computes the temporal collaboration vector $Z_i$ of each cluster which
represents the association weight between the entities in a cluster and $u$ over the
time axis. Line 5 obtains the $TM$-score using temporal mobility model that we discuss in
Section~\ref{sec:tm}. Line 6 tunes the model parameter $\alpha$ for the given
dataset and Line 7 returns the desired score for the node $u$. A high value of
this score is more likely to represent a pure node and a small value makes the
node more likely to be a multi-node. Thus our method can be used as a pre-filter
to identify a small set of nodes that are more likely candidate for being a multi-node.

Given the collaboration network $G$ and a specific vertex $u$ as input,
generation of ego network $G_u (V_u, E_u)$ takes $\mathcal{O}(\lvert V_u \rvert + \lvert
E_u \rvert)$ time. The computational complexity of MCL algorithm is
$\mathcal{O}(t\cdot \lvert V_u \rvert^{3})$ in the worst case, where $t$ is the number of
iterations until convergence. The steps in Line 3 and Line 4 read 
the similarity matrix of $G_u$ using an adjacency list representation; thus the complexity 
of these steps is roughly $\mathcal{O}(\lvert V_u \rvert +\lvert E_u \rvert)$. 
The KL divergence computation in Line 5 uses Equation~\ref{eq:tm-score} to compute
$TM\mbox{-}score$, which has a cost of $\mathcal{O}(cK^{2})$, where $K$ is the
number of clusters after MCL clustering. Thus, the overall time complexity of Algorithm~\ref{alg1} is
$\mathcal{O}(\lvert V_u \rvert^{3})$. However, note that the above 
complexity bound is over the size of the ego network instead of the entire 
collaboration network $G$, which makes the proposed method very efficient.
We also present the running time of our method in ~\ref{sec:runningtime} over
a set of real-life networks.

\subsection{Supervised Classification Setup}\label{sec:centrality based features}
In a supervised classification setting, we can use the $NC$-score and the $TM$-score as
classification features.  For this, we first build a training dataset in which
the exact labels (positive for a multi-node, negative otherwise) of each of the
instances are known.  Then we can use any of our favorite classification
methods to build a model, which can later be used for predicting the label for
an unknown data instance.
Supervised classification setup enables adding of more features for the classification 
task. So, in this setup, we also consider centrality-based graph topology features,
in addition to the $NC$-score and $TM$-score.

We consider four distinct centrality-based features~\cite{Jackson:08}: degree centrality, betweenness
centrality, closeness centrality, and eigenvector centrality. For a given node, $u$
one can always compute the centrality values of $u$ considering the entire collaboration network,
however such a method is not scalable, so we compute $u$'s centrality within its 
ego-network, $G_u$. In $G_u$, $u$ is the central node by construction, but, more
so, if $u$ is a multi-node. This is because when multi-node, $u$ naturally
has a higher than usual degree in $G_u$ leading to a high degree centrality value for $u$.
Also, when $u$ is a multi node, $G_u$ is composed of many disjoint clusters and 
shortest paths among the nodes in distinct clusters must go through $u$, leading to a high 
betweenness centrality for $u$. Similar arguments can also be made for other centrality metrics.
However, one potential issue of computing centrality within $G_u$ is that for different nodes, 
their centrality values in their respective ego networks are not comparable, so we normalize
$u$'s centrality score over the centrality score of all nodes in $G_u$ as below:

\begin{equation}
Centrality\mbox{-}Score(u) = \frac{C(u)}{\sum_{v \in G_u}C(v)};
\end{equation}  
where, $C(x)$ is the centrality value of a node $x$ in $G_u$. We will show in
experimental results that considering centrality metrics as feature improves
the prediction performance.

\section{Experiments and Results}

A key challenge of working on the name entity disambiguation task is to find a
real-life labeled dataset for evaluating the performance of the proposed
solution. There exist real-life collaboration datasets, such as email or phone
networks, but they are anonymized for security concern, so we can't really
obtain the true ambiguity label of the nodes in such networks. Hence, these datasets
are not useful for evaluating the entity disambiguation task. For our experiments, 
we use two well known bibliographic datasets,
DBLP\footnote{http://dblp.org/search/index.php} and
Arnetminer\footnote{http://arnetminer.org}. Both of these datasets are leading
repositories for bibliographic information of computer science conferences and
journals. Also, the ambiguity label of a scientist in any of these networks can be 
determined by manual inspection of the papers published by that scientist.

From both datasets, we select 150 researchers such that half of them
are pure nodes (negative cases), and the rest of them are multi-nodes (positive
cases). We try to make the datasets as representative as possible by choosing a
mix of senior and early career researchers.  To assign the label for a selected
researcher, we manually examine her bibliographic records and also her webpage
profile. Besides, in DBLP, name disambiguation ground truth is already available
for a few of the high profile researchers. We use those ground truths to
double-check our manual labeling. The final dataset is anonymized by mapping
each researcher to a unique id. 

The objective of our experiments with these datasets is to verify whether our
method can distinguish the set of multi-nodes from the pure nodes. For this
validation we use both supervised and unsupervised methodologies. We also
compare our method with the existing state-of-the-art to show that our method
is superior than that both in terms of speed and accuracy.  Besides these, we also
perform experiments to analyze the sensitivity of our method as the parameters
vary.

Our method has only a few parameters, many of which we keep fixed. The first,
$\tau$, denotes the exponential decay rate, which is used while computing the
similarity between a pair of entities in the input network before the
clustering step. We fix the $\tau$ value to $5$ for all of our experiments
since the performance remains stable as we vary $\tau$ for both the datasets.
For the clustering of the collaboration network, we use MCL clustering method.
We use the code provided by the inventor of MCL and set the inflation value of
MCL to $1.4$ (as is recommended by the inventor) for all of our experiments.
For the other data processing, we write our own code using Python.
The experiments are performed on a 2.1 GHz laptop with 4GB memory running Linux
operating system. 

\begin{figure}[h]
\centering
\subfigure[]{
\label{fig:vary-datapoints-arnetminer}
\includegraphics[height=0.45\linewidth ,angle=-90] {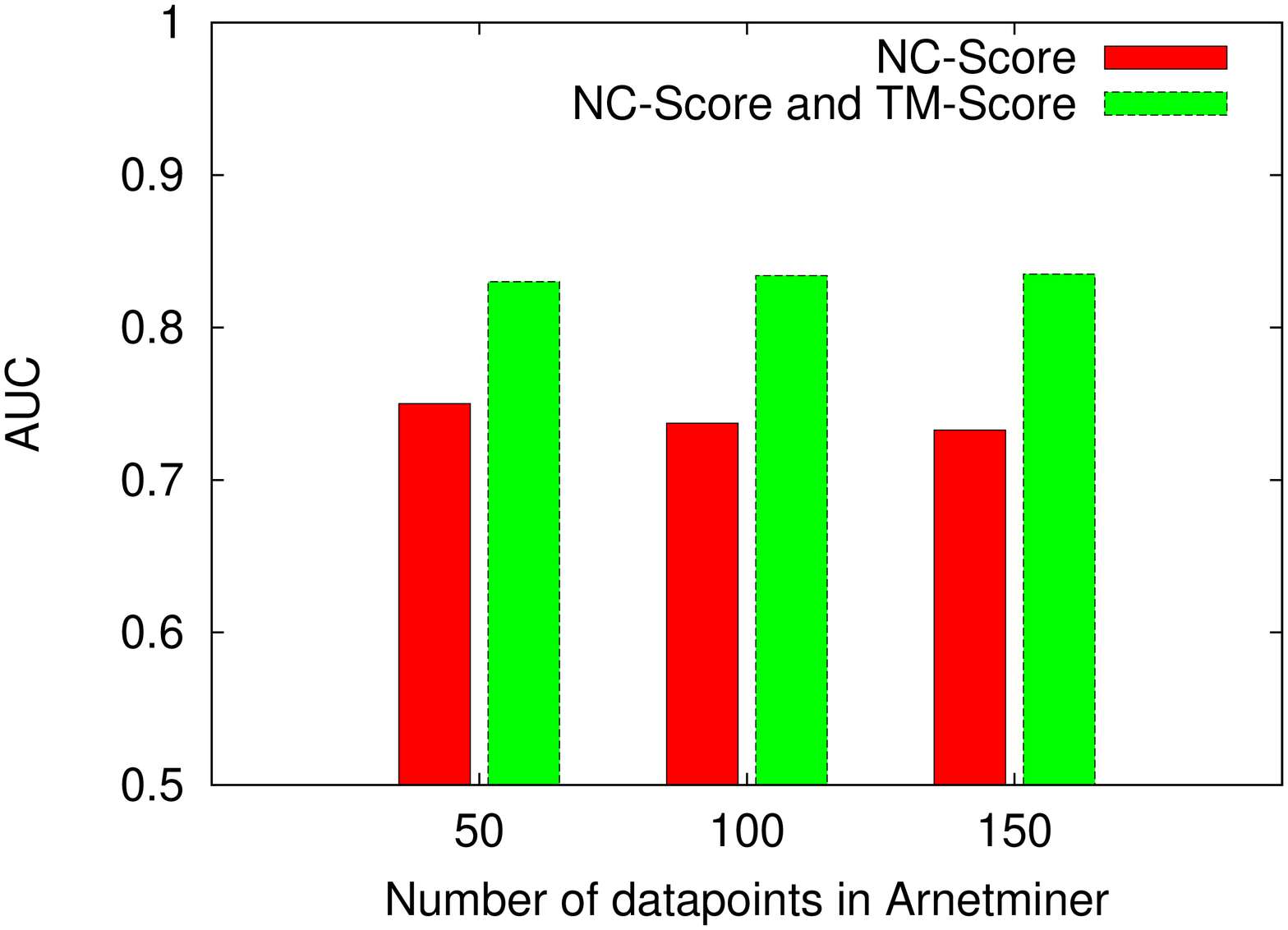}
}
\subfigure[]{
\label{fig:vary-datapoints-dblp}
\includegraphics[height=0.45\linewidth ,angle=-90] {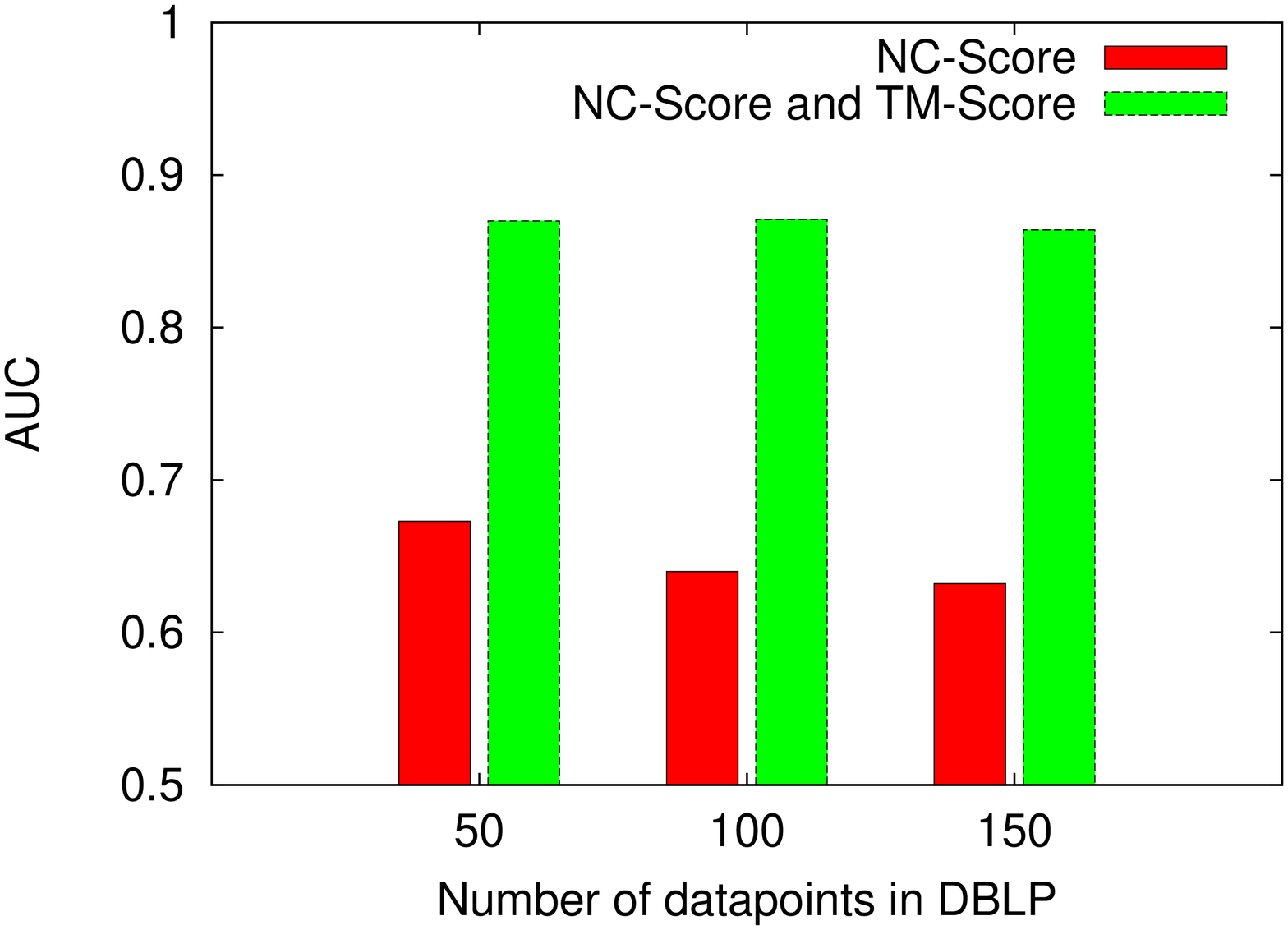}
}
\caption{Unsupervised disambiguation experimental results: (a) on Arnetminer and (b) on DBLP}
\vspace{-0.2in}
\end{figure}

\subsection{Evaluation of Unsupervised Disambiguation}

In unsupervised disambiguation, we do not train a model using a training
dataset, rather we use a linear function (Equation~\ref{eq:s-score}) to obtain
the $s$-score. For evaluating the performance of unsupervised disambiguation,
in this experiment we compute the $s$-score of each of the 150 researchers in
the DBLP and Arnetminer datasets using Equation~\ref{eq:s-score}. We use an
$\alpha$ value of 0.1 for Arnetminer and 0.2 for the DBLP dataset. The
choice of $\alpha$ is fixed by comparing the performance of our method on a
small validation dataset by varying $\alpha$ between 0 and 1.
We use AUC (the area under the ROC curve) as the evaluation
metric of this experiment which we obtain as below.

We sort the $s$-score of the 150 researchers in an increasing order and use
each of the $s$-scores (in that order) as the threshold of our prediction to
obtain a sequence of TPR (true positive rate) and FPR (false positive rate)
pairs. From these (TPR, FPR) datapoints, we draw the ROC curve  and
subsequently compute the AUC value of our
prediction. In this case, the AUC value is essentially the probability that the
$s$-score of a random multi-node (positive data instance) is smaller than
the $s$-score of a random negative data instance. The baseline value for
the AUC is 0.5 and the best value of AUC is 1.  The former case happens if the
$s$-scores of positive and negative instances are non-distinguishable; on the
other hand, the latter case happens when all the $s$-scores of the positive
instances are smaller than all the $s$-scores of the negative instances. For
the DBLP dataset and the Arnetminer dataset, the AUC value that our method
achieves is 0.86 and 0.83 respectively, which, for AUC, is generally considered as excellent.

To show the variance of performance for inputs of different sizes, we construct 2
additional datasets, which are uniformly chosen random subset of the original
dataset. These two datasets have $50$, and $100$ data instances respectively.
For these datasets, we compute the AUC value as we described above. We repeat 
the above random dataset creation process for ten times and compute the
average AUC value for both datasets. We show the AUC comparison
among these datasets in Figure~\ref{fig:vary-datapoints-arnetminer} 
and~\ref{fig:vary-datapoints-dblp} for the cases of Arnetminer and
DBLP, respectively. As we can see for the Arnetminer case, the AUC value
is almost constant (0.83) for all the three
datasets with varying sizes. For DBLP, the datasets with $50$ and $100$ instances
achieve an AUC value of 0.87, whereas the entire dataset with $150$ instances
achieves an AUC value of 0.86.

\begin{table}[h]
\centering
\begin{tabular}{|c|c|c|c|}
\hline
Method & Kernel & DBLP & Arnetminer \tabularnewline
\hline
\multirow{3}{*}{\parbox{2cm}{Our method using NC and TM features}}&Linear & 72.50  & 65.60 \tabularnewline
\cline{2-4}
&Radial basis & 72.31 & 68.82  \tabularnewline
\cline{2-4}
&Sigmoid & 71.90 & 66.20  \tabularnewline
\hline
\hline
\multirow{3}{*}{\parbox{4cm}{Our method using NC, TM and Graph Centrality features}}&Linear & 75.60  & 67.00 \tabularnewline
\cline{2-4}
&Radial basis & 74.71 & 69.42  \tabularnewline
\cline{2-4}
&Sigmoid & 75.30 & 70.03  \tabularnewline
\hline
\hline
\multirow{3}{*}{\parbox{2cm}{Method proposed in~\cite{Hermansson.Kerola.ea:13}}}&GL-3 & 40.67  & 43.62  \tabularnewline
\cline{2-4}
&GL-4 & 41.33  & 44.98  \tabularnewline
\cline{2-4}
&SP & 48.22  & 47.67 \tabularnewline
\hline
\end{tabular}
\caption{Comparison between our method and ~\cite{Hermansson.Kerola.ea:13} using Classification accuracy (\%) on 10-fold cross-validation}
\label{tab:accuracy-result}
\vspace{-0.2in}
\end{table}

\begin{table}[h]
\centering
\begin{tabular}{|c|c|c|c|}
\hline
Method& Kernel type & DBLP & Arnetminer \tabularnewline
\hline
\multirow{3}{*}{\parbox{2cm}{Our method using NC and TM features}}&Linear & 0.80 & 0.76 \tabularnewline
\cline{2-4}
&Radial basis & 0.79 & 0.75 \tabularnewline
\cline{2-4}
&Sigmoid & 0.79 & 0.75 \tabularnewline
\hline
\hline
\multirow{3}{*}{\parbox{4cm}{Our method using NC, TM and Graph Centrality features}}&Linear & 0.83 & 0.80 \tabularnewline
\cline{2-4}
&Radial basis & 0.82 & 0.79 \tabularnewline
\cline{2-4}
&Sigmoid & 0.80 & 0.78 \tabularnewline
\hline
\hline
\multirow{3}{*}{\parbox{2.0cm}{Method proposed in~\cite{Hermansson.Kerola.ea:13}}}&SP & 0.62 & 0.61 \tabularnewline
\cline{2-4}
&GL-3 & 0.63 & 0.62 \tabularnewline
\cline{2-4}
&GL-4 & 0.64 & 0.62 \tabularnewline
\hline
\end{tabular}
\caption{Comparison between our method and ~\cite{Hermansson.Kerola.ea:13} using AUC on 10-fold cross-validation}
\label{tab:AUC-result}
\vspace{-0.2in}
\end{table}

In Figure~\ref{fig:vary-datapoints-arnetminer} and Figure~\ref{fig:vary-datapoints-dblp} 
we also show experimental results that highlight the contribution of $TM$-score in our model. For this we compare the
AUC value that we obtain using TM-score and without using TM-score; the second case
can be obtained by setting $\alpha$=0 in Equation~\ref{eq:s-score}. The AUC value of
these two cases for dataset of different sizes are shown using green (dotted box) and 
red (solid box) bar plots, respectively. As we can see, for both datasets TM-score
significantly improves the AUC score for the cases of all different sizes. For DBLP, the
improvement is particularly significant; for the entire dataset (150 instances), the
AUC without and with TM-score is 0.63 and 0.86 respectively. We guess that the reason 
for such dramatic improvement using TM-score is due to the fact that we 
use academic collaboration datasets, in such a domain temporal mobility occurs rather frequently.

\begin{table}[h]
\centering
\begin{tabular}{|c|c|c|c|c|}
\hline
Method & Kernel Type & Prec & Prec & Prec \\
       &             & @10\%       & @15\%       & @20\% \tabularnewline
\hline
\multirow{2}{*}{\parbox{2cm}{Using NC and TM features}}&Linear & 100 & 100 & 90 \tabularnewline
\cline{2-5}
&Radial basis& 100 & 100 & 90 \tabularnewline
\hline
\hline
\multirow{2}{*}{\parbox{4cm}{Using NC, TM and Graph Centrality features}}&Linear & 100 & 100 & 90 \tabularnewline
\cline{2-5}
&Radial basis& 100 & 100 & 90 \tabularnewline
\hline
\hline
\multirow{3}{*}{\parbox{2.0cm}{Method proposed in~\cite{Hermansson.Kerola.ea:13}}}&SP & 46.7 & 51.7 & 46.7 \tabularnewline
\cline{2-5}
&GL3 & 33.3 & 41.7 & 36.7 \tabularnewline
\cline{2-5}
&GL4 & 46.7 & 41.7 & 44.4 \tabularnewline
\hline
\end{tabular}
\caption{Precision of multi-node class @ Top-k(\%) for DBLP dataset}
\label{tab:precisionpositiveDBLP}
\vspace{-0.2in}
\end{table}

\begin{table}[h]
\centering
\begin{tabular}{|c|c|c|c|c|}
\hline
Method & Kernel Type & Prec & Prec & Prec \\
       &             & @10\%       & @15\%       & @20\% \tabularnewline
\hline
\multirow{2}{*}{\parbox{2cm}{Using NC and TM features}}&Linear & 100 & 100 & 90 \tabularnewline
\cline{2-5}
&Radial basis& 100 & 100 & 90 \tabularnewline
\hline
\hline
\multirow{2}{*}{\parbox{4cm}{Using NC, TM and Graph Centrality features}}&Linear & 100 & 100 & 90 \tabularnewline
\cline{2-5}
&Radial basis& 100 & 100 & 90 \tabularnewline
\hline
\hline
\multirow{3}{*}{\parbox{2cm}{Method proposed in~\cite{Hermansson.Kerola.ea:13}}}&SP & 66.7 & 54.1 & 60.0 \tabularnewline
\cline{2-5}
&GL3 & 60.0 & 58.3 & 56.7 \tabularnewline
\cline{2-5}
&GL4 & 46.7 & 47.5 & 46.7 \tabularnewline
\hline
\end{tabular}
\caption{Precision of multi-node class @ Top-k(\%) for Arnetminer dataset}
\label{tab:precisionpositiveArnetminer}
\vspace{-0.2in}
\end{table}

\subsection{Evaluation of Supervised Disambiguation}\label{sec:sup-dis}
The main objective of our work is to find the $s$-value of a set of
nodes in an unsupervised learning setup. These values can be used for the purpose of pre-filtering 
a small set of suspicious nodes which can be examined more thoroughly in a subsequent stage. However,
we can also use our method in a supervised learning setup to predict whether an entity
is a multi-node or not. For this, we use $NC$-score, $TM$-score, and network centrality based metrics as 
classification features and use SVM classification tool for classification. We use the
LIBSVM library with default parameter setting. During the training phase, we use
the -$b$ option of this library to predict the probability instead of predicting the class label. This
makes it easier to report the performance using AUC metric. While reporting accuracy, we simply predict
the instances with a probability value higher than 0.50 as positive case (a multi-node),
and the remaining as a negative case. 

In Table~\ref{tab:accuracy-result} and 
~\ref{tab:AUC-result}, we show the accuracy and AUC value for both the datasets
using a 10-fold cross validation for various kernels. As we can see, for both 
DBLP and Arnetminer, using these features, the best classification accuracy is achieved 
for the linear kernel and sigmoid kernel, which are 75.60\% and 70.03\%, respectively. For the
case of AUC, all the kernels have almost similar performance, with the best value
of 0.83, and 0.80 for DBLP and Arnetminer, respectively. Considering the fact
that the method works on anonymized network, and only use topological features, accuracy value
around 75\% or AUC value around $0.80$ are indeed commendable.

For this setup, we also report the precision@top-$k$ for $k$ values equal to 10\%, 15\%, and 20\%
of the size of the test datasets. We use 3-fold cross validation for this experiment.
To compute the above precisions, we simply sort the probability 
output of SVM in descending order and find the precision of the model in its desired range.
The results are shown in Table~\ref{tab:precisionpositiveDBLP} and Table~\ref{tab:precisionpositiveArnetminer}
for the two datasets. We see that on DBLP dataset, all the top 15\% of the probability 
values are more than 50\%, thus they are predicted as positive (multi-node) class and
in real-life all those instances also belong to the true positive (multi-node) class, which 
yields a precision of 100\%. For the case of top 20\%, this value drops to 90\%. The result
on the Arnetminer dataset is also similar (see Table~\ref{tab:precisionpositiveArnetminer}).
This result shows that our method is able to place most of the 
true multi-nodes at the top part of its ranking table, as is desired.

For the supervised setting, we also report the results only based on NC and TM features in 
terms of accuracy, AUC and precision@top-$k$. We can observe that adding centrality based
features improves the results in terms of accuracy and AUC. As 
we can see, for both DBLP and Arnetminer, using only these two features, the best classification
accuracy is achieved for the linear kernel and radial basis kernel, which are 72.50\% and 68.82\%,
respectively. For the case of AUC, all the kernels have almost similar performance, with the
best value of 0.80, and 0.76 for DBLP and Arnetminer, respectively. For precision@top-$k$ setup,
the results of using NC and TM as classification features are almost the same 
compared with the results of adding centrality based graph topological features. Overall, 
the marginal improvement of using centrality based features are not that significant, 
which confirms that the $NC$-score, and the $TM$-score that we build are strong features for
this classification task. 

\subsection{Comparison with existing works}\label{sec:comparisonwork}
The work by Hermansson et al.~\cite{Hermansson.Kerola.ea:13} is closely related
to our work as they design a collection of graph kernels to classify
multi-nodes in a supervised learning setup. Their kernels use only the graph
topology, such as, graphlet counts and shortest paths, so they can be used in
an anonymized network for entity disambiguation. To compare with their method,
we run LIBSVM on our dataset using their best performing kernels, namely,
size-3 graphlets (GL3), size-4 graphlets (GL4) and shortest path (SP) kernels.
The kernel values are obtained by source code supplied by the authors. In
Table~\ref{tab:accuracy-result}, Table~\ref{tab:AUC-result},
Table~\ref{tab:precisionpositiveDBLP} and
Table~\ref{tab:precisionpositiveArnetminer}, we compare the performance of our
method that uses $NC$-score, $TM$-score, and centrality based graph topology as 
features with their method that uses topology based kernels, on all three performance metrics, 
accuracy, AUC, and precision@top-$k$. As we can see our method performs much better than their
method on these datasets. For instance, their best kernel achieves only 48.22\%
accuracy on DBLP and 47.67\% accuracy on Arnetminer, whereas our method
achieves 75.60\% and 70.03\% accuracy on DBLP, and Arnetminer. Cross-validation
$t$-test shows that our method is significantly better ($p$-value 0.0051 for
DBLP, and 0.0013 for Arnetminer).
On AUC measure, our method obtains 0.83 an 0.80 on these datasets, whereas their method achieves
a value of 0.64, and 0.62, which are much lower. 

Besides improved performance, another advantage of our method over the methods in~\cite{Hermansson.Kerola.ea:13}
is the superior running time. For a given node in the graph, the running time 
to compute the value of our features are only a few seconds, whereas computing graphlet
kernel values is costly. In our experiments, for some of the nodes, the Matlab
code provided by the authors of~\cite{Hermansson.Kerola.ea:13} took more than 2 days
in a commodity PC.

\begin{figure}[h]
\centering
\subfigure[Parameter Sensitivity of $\tau$ in Arnetminer and DBLP]
{
\label{fig:vary-tau}
\includegraphics[height=0.45\linewidth ,angle=-90] {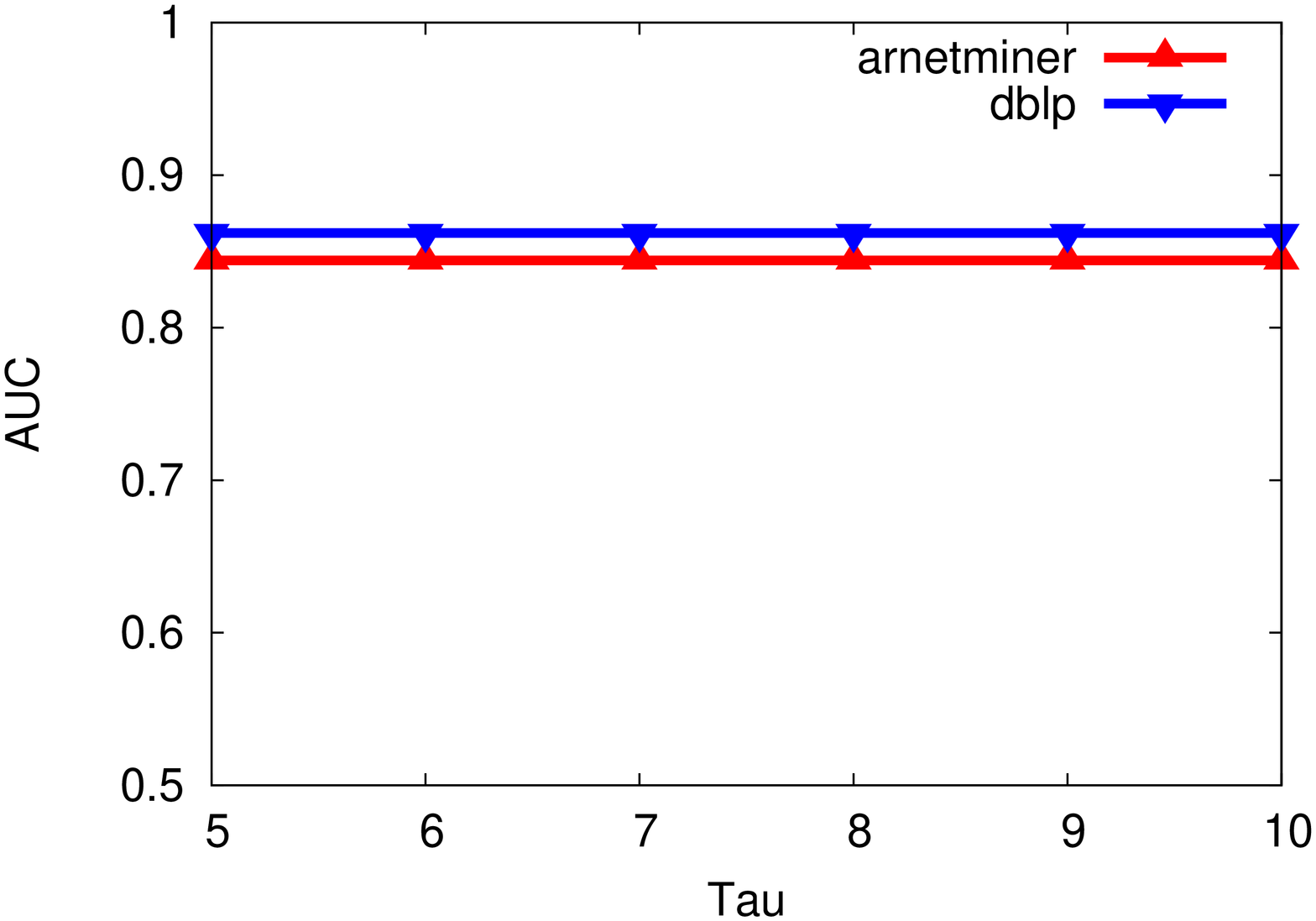}
}
\subfigure[Parameter Sensitivity of $\alpha$ in Arnetminer and DBLP]
{
\label{fig:vary-alpha}
\includegraphics[height=0.45\linewidth ,angle=-90] {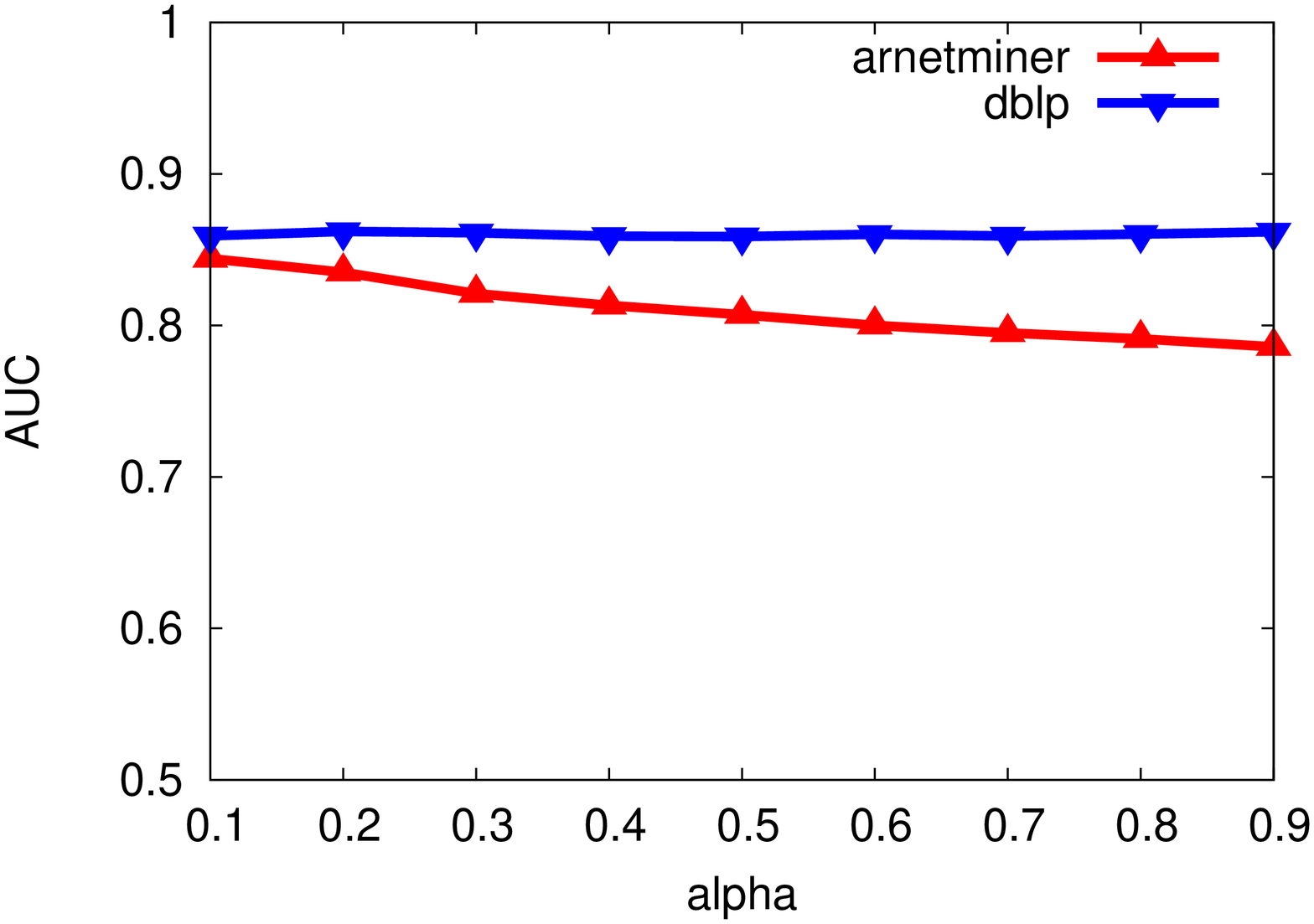}
}
\caption{Parameter Sensitivity Of Our Method On Two Datasets}
\vspace{-0.2in}
\end{figure}

%
%
%
\subsection{Study of Parameter Sensitivity}\label{sec:parametersensitivity}
In case of unsupervised disambiguation task, our method uses two parameters that we set manually, 
one is the exponential decay rate ($\tau$) for similarity computation, and the other is $\alpha$
value in Equation~\ref{eq:s-score}. In this experiment, we see how the performance of the model
changes as we vary the value of these parameters. The result of this experiment is shown in
Figure~\ref{fig:vary-tau} and Figure~\ref{fig:vary-alpha}, where we plot the AUC value for
a range of parameter values. From Figure~\ref{fig:vary-tau} we see that the performance is very 
stable as we vary $\tau$. However, the performance degrades for the choice of $\alpha$ but not
that significantly.

\subsection{Study of Dataset Bias}\label{sec:datasetbias}
\begin{wrapfigure}[12]{r}{0.5\linewidth}
\vspace {-0.35in}
\includegraphics[height=0.95\linewidth ,angle=-90] {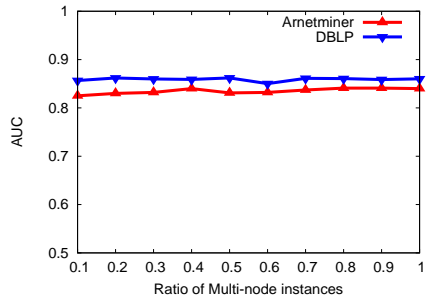}
\caption{Bias Effect in DBLP and Arnetminer}
\vspace{-0.35in}
\label{fig:vary-ratio}
\end{wrapfigure}

Both our datasets are balanced, having equal number of positive and negative cases. 
However in real life scenario it would not be the characteristic of a wild dataset
where the fraction of ambiguous entities is much lower.
In this experiment we change the ratio of these two cases, to find the effect of
dataset bias on the result quality. For this purpose, we change the size of positive instances 
and always keep the negative instances constant which is 75 negative \\ instances during the experiments. 
We randomly select part of positive instances and run the method ten 
times and get the mean AUC as final AUC value. The result
of this experiment is shown in Figure~\ref{fig:vary-ratio}; as we can see 
the performance varies for different ratios but not that significantly.

\begin{table}[h!]
\centering
\begin{tabular}{|c|c|c|}
\hline
Name & Number of  & Running time  \tabularnewline
    & direct neighbors& (seconds) \tabularnewline
\hline
Wei Wang & 1375 & 2.85 \tabularnewline
\hline
Jiawei Han & 523 & 0.55 \tabularnewline
\hline
Philip S. Yu & 488 & 0.45 \tabularnewline
\hline
Wen Gao & 531 & 0.50 \tabularnewline
\hline
Tao Li & 603 & 0.58 \tabularnewline
\hline
\end{tabular}
\caption{Running time result in DBLP}
\label{tab:runningtimeDBLP}
\vspace{-0.2in}
\end{table}

\subsection{Study of Running Time}\label{sec:runningtime}
A very desirable feature of our method is its running time. We have only two features and both
of them can be calculated in a very short time. To compute the running time, we run our unsupervised
disambiguation method on 5 entities from the DBLP datasets that have the largest number of neighbors. 
The running time on these vertices is shown in Table~\ref{tab:runningtimeDBLP}. The Arnetminer dataset
is smaller than the DBLP datasets, so running time on the nodes of this dataset is even smaller.

\begin{table}[h]
\centering
\begin{tabular}{|c|c|c|c|c|}
\hline
Name & Ground      & DBLP & Arnetminer\\ 
     & Truth       & probability & probability\tabularnewline
\hline
Huan Liu & + & 0.80 & 0.68 \tabularnewline
\hline
Tao Li & + & 0.86 & 0.75  \tabularnewline
\hline
Wei Wang & + & 0.87 & 0.77 \tabularnewline
\hline
Tao Xie & + & 0.83 & 0.71  \tabularnewline
\hline
Bin Li & + & 0.86 & 0.75  \tabularnewline
\hline
{\bf Robert Allen} & {\bf +} & {\bf 0.37} & {\bf 0.23} \tabularnewline
\hline
Tim Weninger & - & 3.2e-07 & 0.02 \tabularnewline
\hline
Jianlin Cheng & - & 5.8e-11 & 0.00072 \tabularnewline
\hline
Hector Gonzalez & - & 7.4e-06 & 0.02 \tabularnewline
\hline
{Xifeng Yan} & {-} &{0.38} & {0.42} \tabularnewline
\hline
{\bf Philip S. Yu} & {\bf -} &{\bf 0.80} &{\bf 0.70} \tabularnewline
\hline
\end{tabular}
\caption{Real-life Case Study showing prominent researchers in DBLP and Arnetminer datasets}
\label{tab:probabilitylinearkernel}
\vspace{-0.2in}
\end{table}

\subsection{Real-life Case Study}\label{sec:real-life-case-study}

In Table~\ref{tab:probabilitylinearkernel}, we show the performance of our
method on some of the well-known researchers from data mining and information
retrieval communities. For each of the researchers, we denote the ground truth in
the second column of the table. A positive sign stands for the fact that in
DBLP and Arnetminer datasets the publication records under their names
correspond to more than one real-life entity, and vice-versa. In the same
table we also show the probability value that we obtain by our supervised
disambiguation experiment that we discussed in Section~\ref{sec:sup-dis}. As we
can see for many well known cases of multi-nodes in DBLP, such as Wei Wang,
Huan Liu and Tao Li, our method correctly predicts their labels. A significant
mistake (the mistaken cases are shown in bold fonts) that it makes is that it
also predicts Professor Philip S. Yu to be a multi-node. This is a case of 
false positive, which our method is more susceptible. The reason for it
is that many researchers have multiple disjoint communities that they maintain
concurrently, so for such a researcher the $NC$-score is relatively small;
also since her clusters do not exhibit temporal mobility, the $TM$-score for her case
is also small. So, our method tends to predict such a person as positive. On the
other hand false negative occurs in our method due to the fact that the $TM$-score 
undesirably improves the overall score of a true positive case, even though the $NC$-score of that case is 
very small. One such example is Robert Allen as we show in this table.

\section{Discussion and Conclusion}

In this paper, we propose a novel solution to the entity disambiguation task in
an anonymized network. We discuss the motivation of this task and show that our
solution is useful for solving the entity disambiguation task in a constrained
setting, where biographical features of the actors are not available. We also
discuss how our solution can be used to find a small set of suspects for whom
more detailed analysis can be made in a follow-through process. Another key
strength of our method is that it is robust and it uses a simple model having
only two features, normalized-cut score and temporal mobility score.
Nevertheless, experiments on academic collaboration networks show that our
method have excellent performance. Interestingly, for these datasets, temporal
mobility score improves the prediction performance significantly. We believe
that the dramatic improvement using temporal mobility feature on these datasets 
is due to the fact that in academic domain temporal mobility occurs rather frequently. 
However, due to the unavailability of ground
truth datasets, we could not study whether this phenomenon presents is other
networks, such as Phone call or online social networks, like Facebook.  So we
do acknowledge that the validity of our current work is particularly linked to
academic collaboration networks and we leave the generalization of this work to
networks from other domains as a future research direction.

\bibliographystyle{spmpsci}
\bibliography{namedisambiguation_mod.bib}

\end{document}